\documentclass[aps,twocolumn,prb,amsmath,amssymb,superscriptaddress,longbibliography]{revtex4-2}
\usepackage{url}
\usepackage{graphicx}
\usepackage{color}
\usepackage{natbib}
\usepackage{hyperref}
\usepackage{notes2bib}
\usepackage{mathbbol}
\usepackage{float}
\usepackage{tikz}
\usepackage{xcolor}
\usepackage{dsfont}
\usepackage{svg}
\usepackage[normalem]{ulem}

\begin{document}

\title{Heat current and noise in a point-like junction between normal metal\\ and fractional quantum hall edge state with $\nu = 1/3$}
\author{Redouane Zamoum}
\email{r.zamoum@univ-bouira.dz}
\affiliation{Laboratory of Materials Physics and Optoelectronic Components, Faculty of Exact Sciences, University of Bouira, Algeria}
\author{Houssam Eddine Farhi}
\email{farhihoussameddine@gmail.com}
\affiliation{Department of physics, Faculty of Exact Sciences, University of Bouira, Algeria}
\author{Abdellah Touati}
\email{touati.abph@gmail.com}
\affiliation{Department of physics, Faculty of Exact Sciences, University of Bouira, Algeria}
\author{Adeline Cr\'epieux}
\affiliation{Aix Marseille Univ, Universit\'e de Toulon, CNRS, CPT, Marseille, France}
\date{\today}


\begin{abstract}
In this study, we investigate heat current correlations between two edge states coupled through a constriction in the fractional quantum Hall regime at filling factor $\nu = 1/3$. This configuration allows us to map the system onto an interacting Tomonaga–Luttinger liquid with a local impurity, where the interaction parameter takes the value $\kappa = 1/2$. This specific parameter value enables the application of the refermionization method. By solving the refermionized Hamiltonian, we compute the heat current and evaluate the finite-frequency heat auto- and cross-correlation functions. Our results reveal asymmetric behavior with frequency, corresponding to absorption and emission noise, respectively. Furthermore, we examine several limiting cases, including low- and high-temperature limits as well as strong- and weak-backscattering regimes. In the weak backscattering regime, the Lorenz number approaches $4L_0$ at high voltage, suggesting a rescaling of the Wiedemann–Franz law due to interactions. A detailed study of the heat current fluctuations has made it possible to identify and derive expressions for the dominant contribution in the various regimes of interest,  and to reveal the predominance of partition noise.
\end{abstract}

\maketitle



\section{Introduction}

Quantum point contacts (QPCs) are essential building blocks for nano-electronic devices. Different types of QPCs are realised, such as three-dimensional metallic QPCs obtained from stretching out a metallic wire until breaking \cite{untiedt2004}. Another type is reached from the contact between an STM and a metallic surface, which forms a nanoconstriction \cite{gimzewski1987}. A two-dimensional QPC can be obtained from applying a constriction to a two dimensional electron gas. To confine electrons, a semiconductor in the fractional quantum Hall regime (FQH) can be used, or alternatively a heterostructure of two different semiconductors like AlGaAs-GaAs \cite{beenaker1991}.

Strongly correlated systems like FQH edge channels adopt a QPC geometry. In this case, the QPC plays the role of a scattering center in which the electrons are reflected or transmitted with tunable probability \cite{granger2009,altimiras2010,altimiras2012}. For such a system, the quantization of the thermal conductance is observed for the integer Hall effect \cite{jezouin2013}, as well as for the FQH effect \cite{kane1997,banerjee2017,banerjee2018}. In the same context of QPC geometry, electrical and heat transport properties are investigated. Electrical noise and fractional charge are calculated for junction between different FQH states \cite{sandler1999}. Moreover, in the FQH regime, photo-assisted current and shot noise \cite{crepieux2004}, and heat and mixed photo-assisted noise are computed \cite{ronettientropy2019}. In the case of a Hong-Ou-Mandel interferometer implemented in a quantum Hall bar Laughlin sequence, the charge and heat noise are investigated \cite{ronettiPRB2019}.

Most of the theoretical works on the FQH in the QPC geometry focus on the charge transport, and often disregard thermal transport, even though it encodes crucial information about energy dissipation and the nature of excitation \cite{zhizhou2016}. Among these properties, one can quote: the signature of a crossover between Coulomb-Blockade regime and Kondo effect \cite{ridley2019}, or the full reconstruction of single-electron wave function \cite{moskatles2017}. Heat correlations have been treated in some works, using the Landauer-B\"uttiker formalism in the non interacting case \cite{michelini2014,crepieux2021}. In addition, heat noise was investigated for single electron sources \cite{battista2013,battista2014}. Furthermore, the nature of the minimal excitation for heat transport in driven FQH systems has been shown to be levitons \cite{vannucci2017}. On the experimental side, a protocol has been proposed to measure heat-current noise by frequency dependent fluctuations in temperature and potential \cite{dashti2018}. Besides investigating in the FQH regime, several theoretical works treat about heat-current correlations for various systems using different techniques, such as: molecular junctions \cite{fei2011,wu2009}, quantum dots \cite{zhizhou2016,sanchez2012,eymeoud2015}, double quantum dots \cite{zhizhou2016,agarwalla2015}, quantum wires \cite{krive2001,agarwalla2012}, and qubits \cite{wang2017}. Experimentally, temperature fluctuations in quantum detector have been measured \cite{karimi2020}.

In a more general framework, thermoelectric transport has been investigated through other interesting quantities rather than heat correlations. Let us quote the heat-charge mixed noise, studied for different systems like QPCs or quantum dots \cite{michelini2014,eymeoud2015,michelini2016,eymeoud2016}. Thermal conductance is important to understand thermoelectrical transport\cite{touati2026}. It has been investigated for various systems such as: Tomonaga-L\"uttinger liquids (TLLs) \cite{kane1996}, graphene nanoribbons \cite{hu2011}, electronic fluids \cite{gutman2023}, parallel coupled double quantum dots \cite{zhang2021}, quantum dots with ferromagnetic electrodes \cite{liu2019}, and in a two-channel Kondo system \cite{dalum2020}. Intriguingly, the heat conductance is surprisingly found to be negative for some systems \cite{hu2011,gutman2023,zhang2021,liu2019}. This result arises from the non linear nature of the transport. The negativity of the thermal conductance requires the presence of temperature dependent thermal contact resistance in the junction \cite{hu2013}.

In this work we study a hybrid edge system composed of two chiral channels with different filling factors: one at $\nu =1$, which is a normal metal described by a Fermi liquid; the other has a filling factor $\nu = 1/3$ representing an edge state in the FQH regime described as a TLL. This corresponds to an FQH edge state coupled to a metallic lead through a QPC. To make the model more analytically tractable, the authors in Ref.~\onlinecite{sandler1999} implemented a mapping to completely decoupled TLLs with interaction parameter $\kappa=1/2$. The first TLL is without an impurity (free) and the second contains a localized impurity. The special value of the interaction parameter $\kappa=1/2$ allows us to use bosonization and refermionization techniques \cite{vondelft1998,chamon1996}. Using the new fermionic wave functions we compute exact expressions for the heat current, the thermal conductance, and the non-symmetrized finite frequency heat (NSFFH) noise. Notice that only electronic contribution to the NSFFH correlations is considered.

The paper is organized as follows: in Sec.~II, we describe the system and give a brief overview of the model adopted in this work. In Sec.~III, we present the results for the heat current and thermal conductance. In Sec.~IV, the calculation of the NSFFH correlators is explained, and their expressions in various limit cases are given, such as at equilibrium, in the weak backscattering (WBS) or strong backscattering (SBS) regime, and in the zero temperature limit. We discuss the results in Sec.~V, and conclude in Sec.~VI.



\section{System description and Model}

\begin{figure}[!h]
	\begin{center}
		\includegraphics[width=0.5\textwidth]{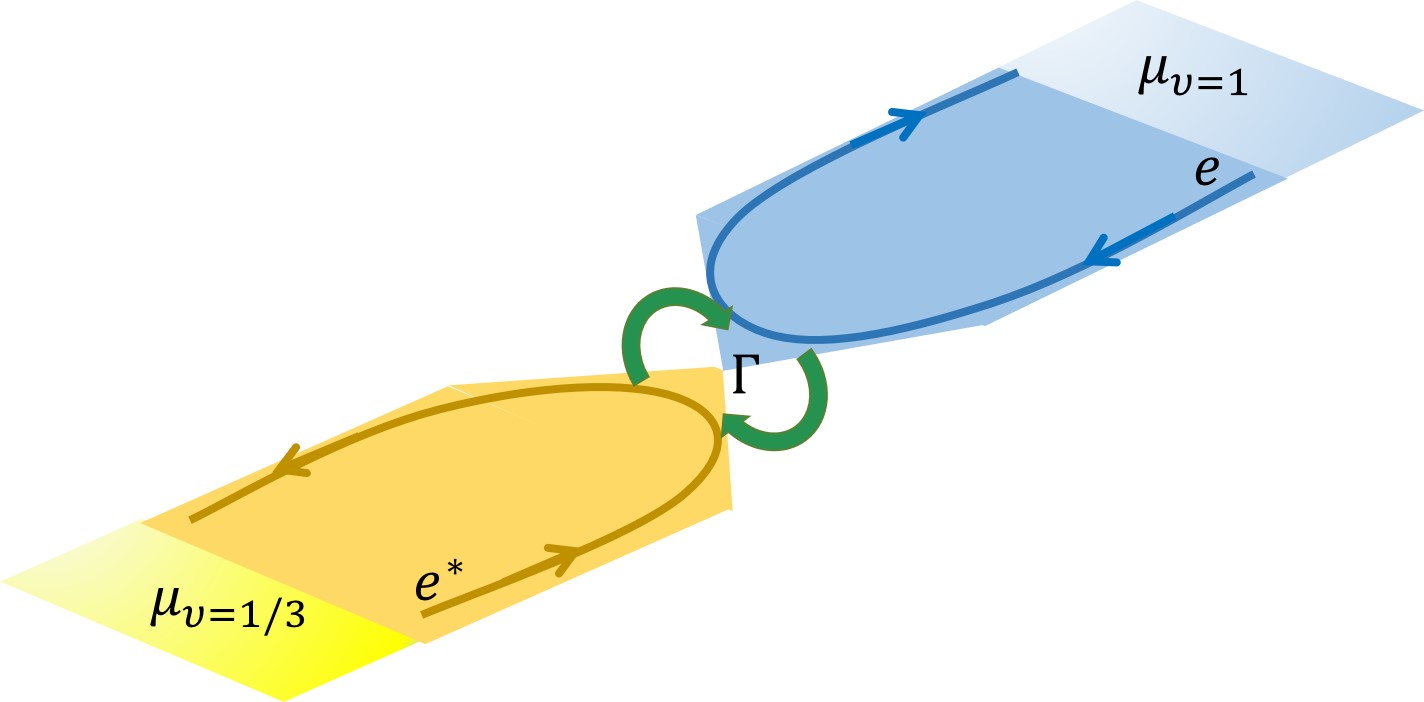}
		\caption{Schematic description of the system representing a metallic lead with chemical potential $\mu_{\nu=1}$ associated with a non-interacting edge state, connected via a QPC to an FQH system with chemical potential $\mu_{\nu=1/3}$ associated with an edge state with filling factor $\nu=1/3$. Electrons $e$ and quasi-particles $e^{\ast}$ can be transmitted from one edge to the other with scattering amplitude $\Gamma$.}\label{fig:model}
	\end{center} 
\end{figure}

We consider a system composed of two chiral edge states, a Laughlin state at filling factor $\nu=1/3$ and a non-interacting state at $\nu=1$ (see Fig.~\ref{fig:model}). The two edges are connected through a QPC, assumed to be a narrow tunable constriction allowing controlled tunneling of electrons or quasi-particles. In the vicinity of the QPC, a particle approaching from one edge can either be transmitted to the opposite edge or reflected back along its original path, thereby introducing both forward and backward scattering channels. In order to describe the system and the associated phenomenon, let $\phi_{\alpha}(x,\tau)$ and $\phi_{\beta}(x,\tau)$ denote the bosonic fields associated with the Laughlin and the non-interacting edges respectively. Each part of the system is modeled by a free chiral Hamiltonian density $\mathcal{H}_{\alpha,\beta}=\frac{v_F}{4\pi}[\partial_x \phi_{\alpha,\beta}(x,\tau)]^2$\: \cite{sandler1999,wen1990}. The tunneling interaction between the two edges is localized at $x=0$, described by a tunneling Hamiltonian density which reads \cite{sandler1999,wen1990}:
\begin{equation}
	\mathcal{H}_{tun}=\Gamma \delta(x) e^{-i\omega_0t}e^{i\left(\frac{1}{\sqrt{\nu}}\phi_{\alpha}(x,\tau)-\phi_{\beta}(x,\tau)\right)}+h.c~.
\end{equation}
The potential difference $V$ between the two sides of the QPC enters a phase shift into the complex tunneling amplitude $\Gamma$, which becomes $\Gamma e^{-i\omega_0t}$. The frequency $\omega_0=e^{\ast}V/\hbar$ can be identified as a Josephson frequency, with $e^{\ast}=e$ for electrons and $e^{\ast}=\nu e$ for quasi-particles. Moreover, the external voltage $e^*V$ can be seen as the difference between the chemical potentials of the Laughlin and non-interacting states ($e^*V=\mu_{\nu=1}-\mu_{\nu=1/3}$).

To make the model easily tractable, a mapping is needed to transform the whole system to a TLL with a  localized impurity. Two steps are necessary: the first one consists of a linear rotation which maps the fields $(\phi_{\alpha},\phi_{\beta})$ into the new ones $(\phi'_{\alpha},\phi'_{\beta})$ without changing the form of the total Hamiltonian density \cite{chamon1997,sandler1998}. However, the new fields $\phi'_{\alpha}$ and $\phi'_{\beta}$ evolve within edge states of an FQH system with an effective filling factor $\nu'$ satisfying $\nu'^{-1}=(1+\nu^{-1})/2$. The particular value of the filling factor $\nu=1/3$ gives the value $\nu'=1/2$ for the effective one. The second step introduces new fields $\phi_+$ and $\phi_-$ through an orthogonal diagonalization of the Hamiltonian density $\phi_{\pm}=(1/\sqrt{2})\big(\phi'_{\alpha} \pm \phi'_{\beta} \big)$. This transformation decouples completely the system into two independent sectors. The field $\phi_+$ describes a free chiral mode unaffected by the tunneling, while $\phi_-$ captures the tunneling interaction part of the system by containing it in a localized point-like impurity at $x=0$. The two sectors represent TLLs with interaction parameter $\kappa=1/2$. The corresponding bosonized Hamiltonian density reads
\begin{equation}\label{hambos}
	\mathcal{H}=\frac{v_F}{4\pi}\Big[ (\partial_x \phi_+)^2+(\partial_x \phi_-)^2 \Big]+\Gamma \delta(x) e^{-i\omega_0t+i\phi_-(x)}+h.c~.
\end{equation}

The actual form of the Hamiltonian density is more suitable for computing observables such as current, thermal conductance, and NSFFH correlations. The special value of the interaction parameter $\kappa=1/2$ allows to employ a refermionization procedure by introducing the new fermions $\psi_{\pm}=(1/\sqrt{2})\mathrm{f} e^{i\phi_{\pm}}$, where $\mathrm{f}$ is a Majorana fermion satisfying $\{ \mathrm{f},\mathrm{f} \}=2$ and $\{ \mathrm{f},\psi_{\pm} \}=0$ \cite{chamon1996}. This makes the Hamiltonian density of Eq.~(\ref{hambos}) exactly solvable, and one can perform a non-perturbative analysis of heat transport properties. The new independent chiral fermion $\phi_-$ owns frequency-dependent transmission and reflection amplitudes given by the following expressions
\begin{equation}\label{coeff}
	t(\omega)=\frac{\omega}{ \omega+i\Gamma/2}~~,~~r(\omega)=\frac{i\Gamma/2}{ \omega+i\Gamma/2}~.
\end{equation}
These expressions fully describe the scattering behavior of the chiral fermions through the QPC.  We set $\hbar=1$ in the following to lighten the formulas.


\section{Heat current and thermal conductance}

The model introduced in the previous section describes two chiral edge modes on which quasiparticles can propagate and can be scattered at a QPC controlled by gate voltage. This corresponds to an FQH system with filling factor $\nu'=1/2$ including two incoming branches and two outgoing branches in the QPC geometry. The two decoupled modes propagate independently, the fields $\psi_+$ are neither affected by scattering nor by applied voltage, and verify $\langle \psi_{+}^{\dag}(x,\tau)\psi_+(x,\tau) \rangle =0$ \: \cite{zamoum2012}. It does not contribute to heat current. On the contrary, the fields $\psi_-$ are affected by scattering and applied voltage, and they are described by the refermionized Hamiltonian density
\begin{eqnarray}\label{hamilton3}
	\mathcal{H}_-(x,\tau) &=&\psi^\dagger_-(x,\tau)\left(-iv_F\frac{\partial}{\partial x}-\omega_0\right)\psi_-(x,\tau)\notag\\
	&+&\sqrt{2\pi}\delta(x)\bigg(\Gamma \psi_-(x,\tau) \mathrm{f}+\Gamma^* \mathrm{f} \psi_-^\dagger(x,\tau)\bigg)~.
\end{eqnarray}
The fields $\psi_-$ completely diagonalize  the Hamiltonian $\mathcal{H}_-$, and the solution of the related equation of motion gives the explicit form of the fields $\psi_-$. We obtain $\psi_-(x<0,\tau)=\psi^\text{in}_-(x,\tau)$ and  $\psi_-(x\ge 0,\tau)=\psi^\text{out}_-(x,\tau)$, with\cite{chamon1996}
\begin{eqnarray}
		\psi^\text{in}_-(x,\tau)&=&\frac{1}{\sqrt{2\pi v_F}}\int_{-\infty}^{+\infty} a_\omega e^{i(\omega+\omega_0)\frac{x}{v_F}}e^{-i\omega\,\tau}d\omega,\label{psi}\label{eq:psi1}\\
		\psi^\text{out}_-(x,\tau)&=&\frac{1}{\sqrt{2\pi v_F}}\int_{-\infty}^{+\infty}  b_\omega e^{i(\omega+\omega_0)\frac{x}{v_F}}e^{-i\omega\,\tau}d\omega,\label{eq:psi2}
\end{eqnarray}
where the operator $b_\omega=t(\omega)a_\omega+r(\omega)a^\dag_{-\omega}$ is a combination of the annihilation and creation operators, $a_\omega$ and $a^\dag_\omega$, which obey the commutation relation $\{a_\omega,a^\dag_{\omega'}\}=\delta_{\omega,\omega'}$. The transmission $t(\omega)$ and the reflection $r(\omega)$ amplitudes are given by Eq.~(\ref{coeff}).

\begin{figure}[!h]
	\begin{center}
		\includegraphics[width=0.4\textwidth]{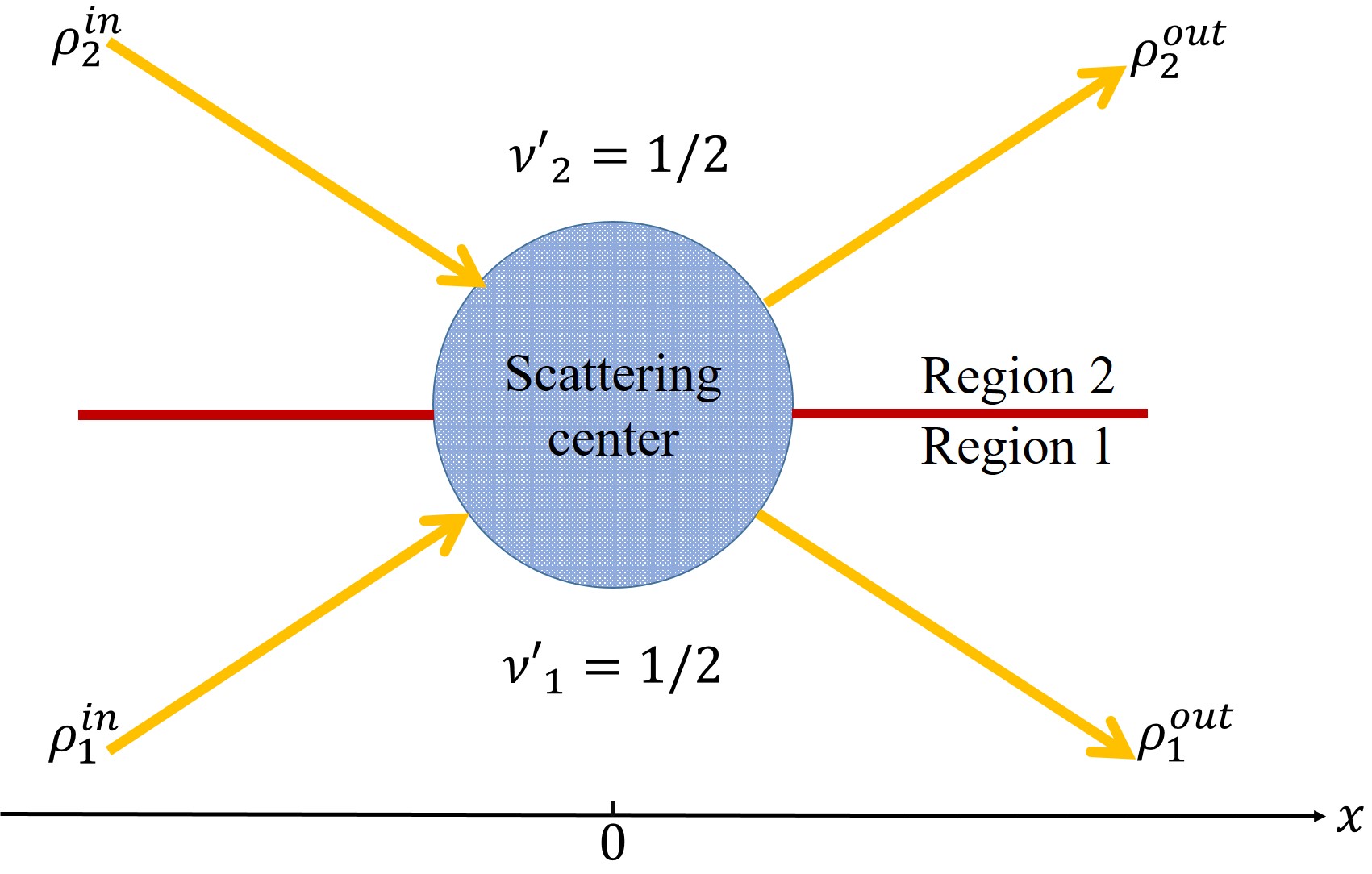}
		\caption{Four probe geometry of the effective system (at $\nu'=1/2$). The incoming densities are at $x<0$ whereas the outgoing densities are at $x>0$.}\label{fig:model2}
	\end{center}
\end{figure}



In this representation, the heat current operators along the edges $i=1,2$ are given by\cite{sandler1999}
\begin{eqnarray}\label{defI1}
 \widehat I_1^{\,h}(x,\tau) =v_F  (\rho_1^\text{in}(x,\tau)-\rho_1^\text{out}(x,\tau)),\\
 \label{defI2}
  \widehat I_2^{\,h}(x,\tau) =v_F  (\rho_2^\text{out}(x,\tau)-\rho_2^\text{in}(x,\tau)),
\end{eqnarray}
where $\rho_i^\alpha$, with $\alpha=\text{in}$ or $\alpha=\text{out}$, and $i=1$ or 2, are the heat density operators associated with the four-probe geometry of the effective system (see Fig.~\ref{fig:model2}). They are defined as
\begin{eqnarray}\label{defrho1}
 \rho_1^\alpha(x,\tau)=\frac{1}{2}\left(\widetilde\rho_{1+}^{\,\alpha}(x,\tau)-\text{sgn}(x)\widetilde\rho_{1-}^{\,\alpha}(x,\tau)\right),\\\label{defrho2}
  \rho_2^\alpha(x,\tau)=\frac{1}{2}\left(\widetilde\rho_{2+}^{\,\alpha}(x,\tau)+\text{sgn}(x)\widetilde\rho_{2-}^{\,\alpha}(x,\tau)\right),
\end{eqnarray}
with $\widetilde\rho_{i\pm}^{\,\alpha}(x,\tau)=\mathcal{H}_\pm^\alpha(x,\tau) -\mu_i\, \mathcal{N}_\pm^\alpha(x,\tau)$ and $\mathcal{N}_\pm^\alpha(x,\tau) =\psi_\pm^{\dag \alpha}(x,\tau) \psi_\pm^\alpha(x,\tau) $, the particle number density operator. The subtraction of $\mu_i\, \mathcal{N}_\pm^\alpha$ corresponds to referencing the density energy with respect to the chemical potential $\mu_i$. We have $\mu_1=eV/2$, $\mu_2=-eV/2$.

The average heat current flowing through an edge of the system is $I_i^h=\langle \widehat I_i^{\,h}(x,\tau) \rangle$. It captures the net flow of heat along the edge $i$. Substituting the expression of the heat density in the heat current definition, using Eqs.~\eqref{eq:psi1} and \eqref{eq:psi2}, and taking the normal order \cite{vondelft1998lett}, we obtain (see Appendix \ref{app1} for details)
	\begin{align}\label{thermalHcurrent1}
		I^h_{i}&=\frac{1}{4\pi}\int_{-\infty}^{+\infty}d\omega(\omega-\mu_{i})\notag\\
		&\times\Big\{
		\big(1+\mathcal{T}(\omega)\big)\big[f_-(\omega)- f_0(\omega)\big]\notag\\
		&+\big(1-\mathcal{T}(\omega)\big)\big[f_+(\omega)-f_0(\omega)\big]\Big\},
	\end{align}
where $f_0(\omega)=1/(1+\exp(\omega/k_BT))$ is the Fermi-Dirac distribution, $f_\pm(\omega)=f_0(\omega\pm eV/2)$, and $ \mathcal{T}(\omega)=\vert t(\omega) \vert^2$ is the transmission coefficient which encodes the non-trivial effects of the interaction. The obtained expression of the total heat current is position independent since the average is a conserved quantity. Equation (\ref{thermalHcurrent1}) describes the behavior of the heat current over all voltage and temperature ranges, from WBS regime to SBS regime, where interactions are beyond the mean-field approach since the system is strongly correlated.

\begin{figure}[!h]
	\begin{center}
		\includegraphics[height=3.7cm]{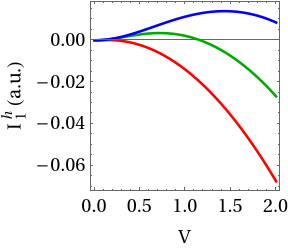}
		\includegraphics[height=3.7cm]{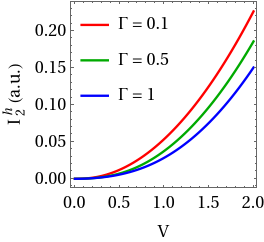}
		\caption{Heat currents $I_1^h$ and $I_2^h$ as a function of voltage for various values of the scattering amplitude $\Gamma$ at temperature $k_BT=0.02$.}\label{fig:Ih1Ih2}
	\end{center} 
\end{figure}

\begin{figure}[!h]
	\begin{center}
		\includegraphics[height=3.8cm]{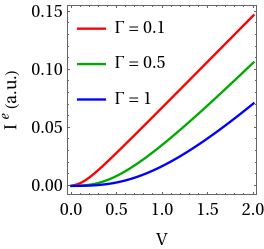}
		\includegraphics[height=3.8cm]{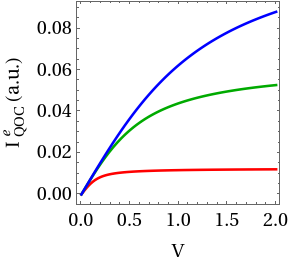}
		\caption{Electrical currents $I^e$ and $I_\text{QPC}^e$ as a function of voltage for various values of the scattering amplitude $\Gamma$ at temperature at $k_BT=0.02$.}\label{fig:IeIeQPC}
	\end{center}
\end{figure}

Figure \ref{fig:Ih1Ih2} shows the heat currents $I_1^h$ and $I_2^h$ as a function of the bias voltage. We see that the variations of $I_1^h$ and $I_2^h$ with $V$ are quite different: $I_1^h$ has a non-monotonic behavior with $V$ at large $\Gamma$, whereas $I_2^h$ follows a power-law. As a first explanation, the power law profile is linked to energy exchange between the system and the environment. In this case, electron transfer is inelastic and the current contains nonlinear contributions at low voltage and low temperature, which represents a Coulomb blockade regime. Such a phenomenon is studied for interacting quantum dots within the Coulomb blockade regime\cite{sierra2015}. In the FQH regime, a zero-bias anomaly in electrical transport for tunnel junction is observed\cite{jiang2010}. The difference between the heat currents corresponds to the Joule effect: $I^h_{2}-I^h_{1}=VI^e$, where $I^e=\frac{e}{4\pi}\int_{-\infty}^{+\infty}d\omega \mathcal{T}(\omega)[f_-(\omega)-f_+(\omega)]$ is the electrical current displayed in the left panel of Fig.~\ref{fig:IeIeQPC}. It shows that the electric current follows  Ohm's law in the WBS regime ($\Gamma\approx 0.1$) and a power law in the SBS regime ($\Gamma\approx 1$).

\begin{figure}[t]
	\begin{center}
		\includegraphics[height=3.7cm]{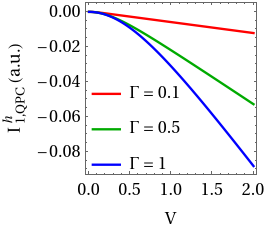}
		\includegraphics[height=3.7cm]{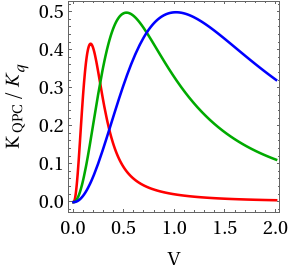}
		\caption{QPC heat current $I^h_{1,\text{QPC}}$  and thermal conductance $K_{\text{QPC}}$ as a function of voltage for various values of~$\Gamma$ at  $k_BT=0.02$.}\label{fig:IhQPCKQPC}
	\end{center}
\end{figure}

Let us now compute the heat current through the QPC, $I^h_{i,\text{QPC}}$, which is defined as the difference between the heat current at perfect transmission, i.e., when $\mathcal{T}(\omega)=1$, and the heat current itself: $I^h_{i,\text{QPC}}=I^h_{i,\mathcal{T}=1}-I^h_{i}$. We find
 \begin{align}\label{rhoout}
	I^h_{i,\text{QPC}}&=\frac{1}{4\pi}\int_{-\infty}^{+\infty} d\omega(\omega-\mu_i)\notag\\
	&\times\big(1-\mathcal{T}(\omega)\big)\big[f_-(\omega)-f_+(\omega)\big],
\end{align}
Although one might initially think by looking at Eq.~(\ref{rhoout}) that $I^h_{1,\text{QPC}}$ and $I^h_{2,\text{QPC}}$ are not related to each other, it turns out that we have $I^h_{2,\text{QPC}}=-I^h_{1,\text{QPC}}$. This can be formally demonstrated by taking into account the fact that $\mathcal{T}(\omega)$ is an even function with respect to the variable $\omega$. This relationship ensures that the total amount of heat in the system remains constant. The left panel of Fig.~\ref{fig:IhQPCKQPC} shows the evolution of $I^h_{1,\text{QPC}}$ as a function of bias voltage for increasing values of $\Gamma$. We observe that contrary to the heat current $I_i^h$ which becomes smaller in absolute value when $\Gamma$ increases, the QPC heat current $I^h_{i,\text{QPC}}$ becomes larger when $\Gamma$ increases.

\begin{figure}[t]
	\begin{center}
		\includegraphics[height=3.9cm]{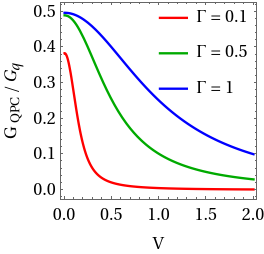}
		\includegraphics[height=3.9cm]{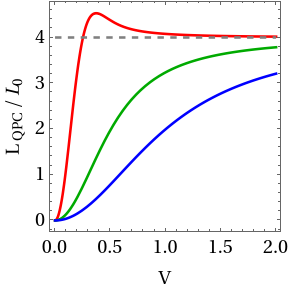}
		\caption{Electrical conductance $G_\text{QPC}$ and Lorenz number $L_\text{QPC}$ as a function of voltage for various values of $\Gamma$ at $k_BT=0.02$.}\label{fig:GQPCLQPC}
	\end{center}
\end{figure}


In order to better understand the heat transport in our system, we study the thermal conductance through the QPC and the Lorenz number in order to see whether the Wiedemann-Franz law applies in our system. The differential thermal conductance is defined as $K_{i,\text{QPC}}={\partial I^h_{i,\text{QPC}}}/{\partial T}$. By starting from Eq.~(\ref{rhoout}), we obtain
	\begin{align}\label{condtherm}
		&K_{i,\text{QPC}}=\frac{3 K_q}{4\pi^3(k_BT)^3 }\int_{-\infty}^{\infty}d\omega(\omega-\mu_i)\big(1-\mathcal{T}(\omega)\big)\notag\\
		&\times\bigg[\Big(\omega+ \frac{eV}{2}\Big)F_{+}(\omega)-\Big(\omega- \frac{eV}{2}\Big)F_{-}(\omega)\bigg],
	\end{align}
where $F_{\pm}(\omega)=f_\pm(\omega)[1-f_\pm(\omega)]$, and $K_q=\pi^2k_B^2T/(3h)$ is the quantum of thermal conductance\cite{jezouin2013}. It can be shown that $K_{2,\text{QPC}}=-K_{1,\text{QPC}}$, and we define $K_{\text{QPC}}=|K_{i,\text{QPC}}|$. The Lorenz number $L_\text{QPC}$ is given by $L_\text{QPC}=K_\text{QPC}/(TG_\text{QPC})$, where $G_\text{QPC}=\partial I^e_\text{QPC}/\partial V$ is the electrical conductance through the QPC, and $I^{e}_\text{QPC}$ is the electrical current through the QPC given by \cite{zamoum2012,zamoum2013}
\begin{align}\label{elec}
	I^{e}_\text{QPC}=\frac{e}{4\pi }\int_{-\infty}^{\infty}d\omega~\big(1-\mathcal{T}(\omega)\big)[f_-(\omega)-f_+(\omega)].
\end{align}
The calculation yields
\begin{align}\label{condelec}
	G_\text{QPC}=\frac{G_q}{4 k_B T}\int_{-\infty}^{\infty}d\omega~\big(1-\mathcal{T}(\omega)\big)\big[F_{+}(\omega)+F_{-}(\omega) \big],
\end{align}
where $G_q=e^2/h$ is the quantum of electrical conductance. The ratio between the quantum of thermal conductance and the quantum of electrical conductance gives the Lorenz factor $L_0=K_q/(TG_q) =\pi^2k_B^2/(3e^2)$.

In the right panel of Fig.~\ref{fig:IhQPCKQPC} is plotted $K_{\text{QPC}}$ in units of $K_q$, as a function of the voltage for different values of scattering amplitude~$\Gamma$. We observe that the QPC thermal conductance is zero at zero voltage, that it reaches a maximum when $eV \approx \Gamma$, and that it converges to zero at large voltage. In Fig.~\ref{fig:GQPCLQPC}, we plot the electrical conductance $G_\text{QPC}$, in units of $G_q$, and the Lorenz ratio $L_\text{QPC}$, in units of~$L_0$. In the SBS regime (blue curve), the electrical conductance reaches its maximum value $G_q/2$ at zero voltage and decreases to zero at large voltage, in agreement with the zero-temperature limit of Eq. (\ref{condelec}), which is $G_\text{QPC}^\text{T=0}=G_q\Gamma^2/(2(e^2V^2+\Gamma^2))$ as shown in Table~\ref{tab:limit_IK}. The Lorenz number is equal to zero at zero voltage, since the QPC thermal conductance is zero and the electrical conductance is non-zero at zero voltage, and it converges to a constant value equal to  $L_0^*=4L_0$ at large voltage in the WBS regime (see the right panel of Fig.~\ref{fig:GQPCLQPC}). This leads us to conclude that the Wiedemann-Franz law is rescaled in our system. Indeed, in the regime $k_BT\ll \Gamma, eV$, the analytical study of the limits of Eq.~(\ref{condtherm}) gives the expression $K_\text{QPC}^\text{T=0}=2K_q e^2V^2\Gamma^2/(e^2V^2+\Gamma^2)^2$, which leads to $L_{\text{QPC}}=4L_0e^2V^2/(e^2V^2+\Gamma^2)$. It reduces to $L_{\text{QPC}}=L_0^*$ at large bias voltage $eV$ compared to the scattering amplitude $\Gamma$, where $L_0^*=\pi^2k_B^2/(3\kappa^2 e^2)$ is an effective Lorenz factor renormalized by interactions, with $\kappa e=e/2$ the quasiparticle charge. This leads to a breakdown of the Wiedemann-Franz law due to the fact that electrical transport and thermal transport are affected differently by interactions\cite{roche2026}.

\begin{table}[h!]
\renewcommand\arraystretch{2.7}
\begin{tabular}{|l||c|c|c|}
\hline
 &  $G_{\text{QPC}}$ & $K_{\text{QPC}}$ & $L_{\text{QPC}}$ \\
 \hline
  \hline
  Low $T$ limit & $\cfrac{G_q\Gamma^2}{2(e^2V^2+\Gamma^2)}$  & $\cfrac{2K_qe^2V^2\Gamma^2}{(e^2V^2+\Gamma^2)^2}$ & $\cfrac{4L_0e^2V^2}{(e^2V^2+\Gamma^2)}$  \\
 \hline
 WBS limit & $\cfrac{G_q\Gamma^2}{2e^2V^2}$   & $\cfrac{2K_q\Gamma^2}{e^2V^2}$  &   $L_0^*=4L_0$  \\
 \hline
  SBS limit & $\cfrac{G_q}{2}$  & $\cfrac{2K_qe^2V^2}{\Gamma^2}$    &  $\cfrac{4L_0e^2V^2}{\Gamma^2}$   \\

 \hline
  High $T$ limit & $\cfrac{\pi G_q\Gamma}{16 k_BT} $ & $\cfrac{3K_qe^2V^2\Gamma}{32\pi k_B^3T^3}$ &   $\cfrac{V^2}{2T^2}$  \\
 \hline
\end{tabular}
\caption{Expressions for electrical conductance, thermal conductance and Lorenz ratio in the various limits of interest.}\label{tab:limit_IK}
\end{table}


\section{Heat noise}


The NSFFH noise associated with the chiral edges is defined as the Fourier transform of the non-symmetrized correlator of heat currents taken at two different times
 \begin{eqnarray}
  \mathcal{S}_{ij}^{\,h}(\omega)=\int d\tau e^{i\omega \tau}\langle \delta \widehat I_i^{\,h}(x,0) \delta\widehat I_j^{\,h}(x',\tau) \rangle,
 \end{eqnarray}
where $\delta \widehat I_i^{\,h}(x,\tau)=\widehat I_i^{\,h}(x,\tau)-\langle  \widehat I_i^{\,h}\rangle$. By using Eqs.~(\ref{defI1}) and (\ref{defI2}), we have
 \begin{eqnarray}\label{relation1}
  \mathcal{S}_{ij}^{\,h}(\omega)=\sum_{\alpha\beta}p_i^\alpha p_j^\beta  \mathcal{S}_{ij}^{\alpha/\beta}(\omega),
  \end{eqnarray}
where $p_1^\text{in}=1$, $p_1^\text{out}=-1$, $p_2^\text{in}=-1$, $p_2^\text{out}=1$, and where the heat density noise is defined as
\begin{eqnarray}
 \mathcal{S}_{ij}^{\alpha/\beta}(\omega)=v_F^2\int_{-\infty}^\infty d\tau e^{i\omega \tau} \langle \delta \rho_i^{\alpha}(x,0) \delta \rho_j^{\beta}(x',\tau) \rangle,
\end{eqnarray}
with $\delta \rho_i^{\alpha}(x,\tau)= \rho_i^{\alpha}(x,\tau)-\langle  \rho_i^{\alpha}\rangle $, and $\rho_i^{\alpha}(x,\tau)$ given by Eqs.~(\ref{defrho1}) and~(\ref{defrho2}). Since the fields $\psi_-$ and $\psi_+$ are decoupled, we have
 \begin{eqnarray}\label{relation2}
\mathcal{S}_{ij}^{\alpha/\beta}(\omega)
 =\mathcal{ \widetilde  S}_{i+j+}^{\alpha/\beta}(\omega)
 +p_i^\alpha p_j^\beta\mathcal{ \widetilde  S}_{i-j-}^{\alpha/\beta}(\omega),
 \end{eqnarray}
where
\begin{eqnarray}
\mathcal{ \widetilde  S}_{i\pm j\pm}^{\alpha/\beta}(\omega)=\frac{v_F^2}{4}\int d\tau e^{i\omega \tau} \langle \delta \widetilde\rho_{i\pm}^{\,\alpha}(x,0) \delta \widetilde\rho_{j\pm}^{\,\beta}(x',\tau) \rangle.
 \end{eqnarray}
 Thus, to get the NSFFH noise, we have to calculate the heat density correlators $\mathcal{ \widetilde  S}_{i\pm j\pm}^{\alpha/\beta}(\omega)$ in the various branches of the system: there are sixteen of them in total. Indeed, the indices $i$ and $j$ can take the values 1 or 2, and the indices $\alpha$ and $\beta$ can take the values ``in'' or~``out'' (see Fig. \ref{fig:model2}).


 The details of the calculation of the correlators $\mathcal{ \widetilde  S}_{i\pm j\pm}^{\alpha/\beta}(\omega)$ are given in Appendix \ref{app2}. For the free field $\psi_+$, the correlator does not depend on $\alpha,\beta$, and is equal to $
\mathcal{ \widetilde  S}_{i+j+}^{\alpha/\beta}(\omega)= \mathcal{ \widetilde  S}_{ij}^{0}(\omega)$, where
  \begin{eqnarray}\label{expSinin}
 \mathcal{ \widetilde  S}_{ij}^{0}(\omega)&=&\frac{1}{16\pi^2}\int_{-\infty}^{+\infty} d\omega'(\omega'-\omega-\mu_i)(\omega'-\mu_j)\notag\\
&&\times f_-(\omega')(1-f_-(\omega'-\omega)).
 \end{eqnarray}
For the field $\psi_-$, the correlator $\mathcal{ \widetilde  S}_{i-j-}^{\alpha/\beta}(\omega)$ does depend on $i,j$ and $\alpha,\beta$. The calculation yields to $
\mathcal{ \widetilde  S}_{i-j-}^{\,\text{in}/\text{in}}(\omega)= \mathcal{ \widetilde  S}_{ij}^{0}(\omega)$, $\mathcal{ \widetilde  S}_{i-j-}^{\,\text{in}/\text{out}}(\omega)=(\mathcal{ \widetilde  S}_{i-j-}^{\,\text{out}/\text{in}}(\omega))^*$, where
 \begin{eqnarray}\label{expSoutin}
 \mathcal{ \widetilde  S}_{i-j-}^{\,\text{out}/\text{in}}(\omega)
 &=&\frac{1}{16\pi^2}\int_{-\infty}^{+\infty}  d\omega' (\omega'-\mu_j)\notag\\
 &&\times[(\omega'-\omega-\mu_i)t^*(\omega') t(\omega'-\omega)\notag\\
 &&+(\omega'+\mu_i)r^*(\omega')r(\omega'-\omega)]\notag\\
 &&\times f_-(\omega')[1-f_-(\omega'-\omega)],
\end{eqnarray}
 and
 \begin{eqnarray}\label{expSoutout}
 \mathcal{ \widetilde  S}_{i-j-}^{\,\text{out}/\text{out}}(\omega)
&=&\frac{1}{16\pi^2}\int_{-\infty}^{+\infty} d\omega'(\omega'-\omega-\mu_i)(\omega'-\mu_j)\notag\\
&&\times[ \mathcal{T}(\omega') f_-(\omega') +\mathcal{R}(\omega')f_+(\omega') ]\notag\\
 &&\times  [ \mathcal{T}(\omega'-\omega)(1- f_-(\omega'-\omega)) \notag\\
&&+\mathcal{R}(\omega'-\omega)(1- f_+(\omega'-\omega))]\notag\\
 &&+\frac{1}{16\pi^2}\int_{-\infty}^{+\infty} d\omega'(\omega'+\mu_i)(\omega'-\mu_j)\notag\\
&&\times t(\omega')r^*(\omega')t(\omega'-\omega)r^*(\omega'-\omega) \notag\\
 &&\times [f_+(\omega')-f_-(\omega')]\notag\\
 &&\times[f_+(\omega'-\omega)- f_-(\omega'-\omega)]
 \notag\\
&&+\frac{\omega_F\Gamma^2}{4} ( f_-(\omega) +f_+(\omega)),
\end{eqnarray}
with $\mathcal{R}(\omega')=1-\mathcal{T}(\omega')$, and $\omega_F=v_F/a$, where $a$ is the cutoff distance of the TLL model. The result $
\mathcal{ \widetilde  S}_{i-j-}^{\,\text{in}/\text{in}}(\omega)= \mathcal{ \widetilde  S}_{ij}^{0}(\omega)$ is related to the fact that the field $\psi_-^\text{in}$ does not feel the scattering: thus, it can be considered as free, like $\psi_+$. The corresponding noise is purely  thermal noise, which vanishes at zero temperature.


To discuss the results obtained for the heat current noise, we need to know the behavior of the correlators $\mathcal{ \widetilde  S}_{i-j-}^{\alpha/\beta}(\omega)$ in the various limits starting from their expressions given by Eqs.~(\ref{expSinin}) to (\ref{expSoutout}). These limits are calculated in Appendix \ref{app3} and summarized in Table~\ref{tab:limit} in the case where $\omega\ge 0$ (emission noise) and $V\ge 0$. They correspond to zero-frequency limit ($\omega=0$), equilibrium limit ($\omega=V=0$), zero-temperature limit ($T=0$), WBS regime or SBS regime. This shows that in the WBS regime, we have $\mathcal{ \widetilde  S}_{i-j-}^{\alpha/\beta}(\omega)=\mathcal{ \widetilde  S}_{ij}^{0}(\omega)$, $\forall \alpha,\beta$, which corresponds to the thermal noise. In the zero-temperature limit, $T=0$, the only non-vanishing correlator is  $ \mathcal{ \widetilde  S}_{i-j-}^{\,\text{out}/\text{out}}(\omega)$. In the SBS regime at  $T=0$, $ \mathcal{ \widetilde  S}_{i-j-}^{\,\text{out}/\text{out}}(\omega)$ is a pure partition heat noise.

\begin{table}[h!]
\renewcommand\arraystretch{2.2}
\begin{tabular}{|l||c|c|c|}
\hline
 &  $ \mathcal{ \widetilde  S}_{i-j-}^{\,\text{in}/\text{in}}(\omega)$ & $ \mathcal{ \widetilde  S}_{i-j-}^{\,\text{out}/\text{in}}(\omega)$ &  $ \mathcal{ \widetilde  S}_{i-j-}^{\,\text{out}/\text{out}}(\omega)$ \\
 \hline
  \hline
$\omega=0$ & $\mathcal{ \widetilde  S}_{ij}^{0}(0)$ & $\ne  \mathcal{ \widetilde  S}_{ij}^{0}(0)$ &   $\ne  \mathcal{ \widetilde  S}_{ij}^{0}(0)$   \\
 \hline
$V=0 $, $\omega=0$ & $\mathcal{ \widetilde  S}_{ij}^{0}(0)|_{V=0}$  & $\mathcal{ \widetilde  S}_{ij}^{0}(0)|_{V=0}$  & $\ne\mathcal{ \widetilde  S}_{ij}^{0}(0)|_{V=0}$    \\
 \hline
  $T=0$ & 0 & 0 & $\ne 0$  \\
 \hline
 WBS ($\Gamma=0$) & $\mathcal{ \widetilde  S}_{ij}^{0}(\omega)$  & $\mathcal{ \widetilde  S}_{ij}^{0}(\omega)$  &  $\mathcal{ \widetilde  S}_{ij}^{0}(\omega)$  \\
 \hline
  SBS (large $\Gamma$) &  $\mathcal{ \widetilde  S}_{ij}^{0}(\omega)$ & $\ne  \mathcal{ \widetilde  S}_{ij}^{0}(\omega)$ &   $\ne  \mathcal{ \widetilde  S}_{ij}^{0}(\omega)$  \\
 \hline
   SBS (large $\Gamma$), $T=0$ & 0 & 0 &   $\cfrac{\omega_F\Gamma^2}{4}f_-(\omega)$  \\
 \hline
\end{tabular}
\caption{Expressions for heat density correlators in the various limits of interest. The results apply for $\omega, V\ge 0$.}\label{tab:limit}
\end{table}


\section{Discussion}

In this section, we discuss the evolution of the QPC heat noise, $ \mathcal{S}_{ij,\text{QPC}}^h(\omega)$, according to the frequency, making a distinction between auto-correlators obtained when $i=j$, and cross-correlators obtained when $i\ne j$. From Eqs.(\ref{relation1}) and (\ref{relation2}), we have
\begin{eqnarray}
\mathcal{S}_{ij}^h(\omega)=\sum_{\alpha\beta} \mathcal{\widetilde S}_{i-j-}^{\,\alpha/\beta}(\omega),
\end{eqnarray}
since the correlators $\mathcal{\widetilde S}_{i+j+}^{\,\alpha/\beta}(\omega)$, which are independent on $\alpha$ and $\beta$, give a vanishing contribution to $\mathcal{S}_{ij}^h(\omega)$. By using the definition of the backscattering current operator, $\widehat I_{i,\text{QPC}}^{\,h}(x,\tau) =v_F  (\rho_i^\text{in}(x,\tau)+\rho_i^\text{out}(x,\tau))$, we show that the QPC heat noise is given by
\begin{eqnarray}\label{QPCnoise}
 \mathcal{S}_{ij,\text{QPC}}^h(\omega)=4\mathcal{\widetilde S}_{ij}^{0}(\omega)+\sum_{\alpha\beta} p_i^{\alpha}p_j^\beta \mathcal{\widetilde S}_{i-j-}^{\,\alpha/\beta}(\omega).
\end{eqnarray}
We remark that since we have $\mathcal{\widetilde S}_{i-j-}^{\,\beta/\alpha}(\omega)=(\mathcal{\widetilde S}_{i-j-}^{\,\alpha/\beta}(\omega))^*$, the heat noises $\mathcal{S}_{ij}^h(\omega)$ and $\mathcal{S}_{ij,\text{QPC}}^h(\omega)$ are real quantities. To highlight the specificity of the heat noise, we also calculate the QPC electrical noise. Its expression is
 \begin{eqnarray}\label{expSe}
 \mathcal{S}_{ij,\text{QPC}}^e(\omega)&=&4\hbar\omega N(\omega)G_q+\frac{(2\delta_{ij}-1)e^2}{8\pi}\int_{-\infty}^{+\infty} d\omega'\notag\\
 &\times&\sum_{s,s'=\pm}\mathcal{M}_{ss'} (\omega',\omega)f_s(\omega')[1-f_{s'}(\omega'-\omega)],\notag\\
\end{eqnarray}
where $ N(\omega)=(\exp(\omega/k_BT)-1)^{-1}$ is the Bose-Einstein distribution function, and
 \begin{eqnarray}
   \mathcal{M}_{+-} (\omega',\omega)&=&|t(\omega')-t(\omega'-\omega)|^2+\mathcal{T}(\omega'-\omega)-\mathcal{T}(\omega'),\notag\\
  \mathcal{M}_{-+} (\omega',\omega)&=&|t(\omega')-t(\omega'-\omega)|^2+\mathcal{T}(\omega')-\mathcal{T}(\omega'-\omega),\notag\\
  \mathcal{M}_{--} (\omega',\omega)&=&|t(\omega')-t(\omega'-\omega)|^2-\mathcal{R}(\omega')-\mathcal{R}(\omega'-\omega),\notag\\
  \mathcal{M}_{++} (\omega',\omega)&=&-\mathcal{M}_{--} (\omega',\omega).
 \end{eqnarray}
The contribution $4\hbar\omega N(\omega)G_q$ to the QPC electrical noise corresponds to the Johnson-Nyquist noise. The expressions of $\mathcal{S}_{ij,\text{QPC}}^h(\omega)$ and $\mathcal{S}_{ij,\text{QPC}}^e(\omega)$ in the various limits of interest are calculated from Table~\ref{tab:limit} and summarized in Table~\ref{tab:limit_SQOC}.

\begin{table}[h!]
\renewcommand\arraystretch{2.}
\begin{tabular}{|l||c|c|c|}
\hline
 &  $\mathcal{S}_{ij,\text{QPC}}^h(\omega)$ & $\mathcal{S}_{ij,\text{QPC}}^e(\omega)$  \\
 \hline
  \hline
  WBS limit ($\omega=0$)  & $\cfrac{k_BTe^2V^2}{4\pi^2\hbar}\;\delta_{i2}\,\delta_{j2}$ & $4k_BTG_q$  \\
 \hline
 WBS limit ($T=0$)  &  $\cfrac{\omega N(\omega)}{4\pi^2}\bigg(e^2V^2-\cfrac{2\hbar^2\omega^2}{3}\bigg)$ & $4\hbar\omega N(\omega)G_q$\\
 \hline
  SBS limit & $(2\delta_{ij}-1)\cfrac{\omega_F\Gamma^2}{4}f_-(\omega)$ & $4\hbar\omega N(\omega)G_q$ \\

 \hline
  High $T$ limit & $4\hbar\omega N(\omega)TK_q$ & $4\hbar\omega N(\omega)G_q$ \\
 \hline
\end{tabular}
\caption{Expressions for QPC heat and electrical noises in the various limits of interest. The results applied for $\omega, V\ge 0$.}\label{tab:limit_SQOC}
\end{table}


\begin{figure}[!h]
	\begin{center}
		\includegraphics[height=3.8cm]{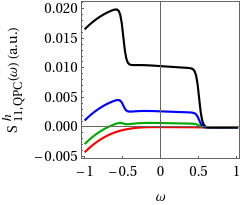}
		\includegraphics[height=3.8cm]{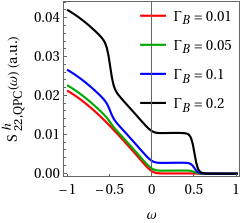}
		\includegraphics[height=3.6cm]{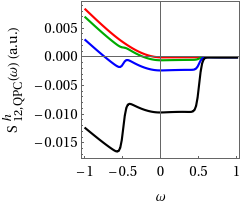}
		\includegraphics[height=3.6cm]{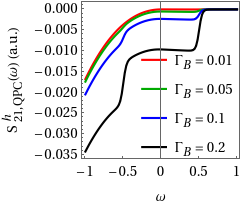}
		\caption{Heat auto-correlators and cross-correlators as a function of frequency for various values of~$\Gamma$. We set $eV=1$ and $k_BT=0.02$.}\label{fig:Sh11QPCSh22QPC}
	\end{center}
\end{figure}

Figure~\ref{fig:Sh11QPCSh22QPC} shows the behavior of the auto-correlators $\mathcal{S}_{ii,\text{QPC}}^h(\omega)$ and cross-correlators $\mathcal{S}_{i\ne j,\text{QPC}}^h(\omega)$ as a function of frequency $\omega$, for various scattering amplitudes~$\Gamma$, at fixed voltage $V$ and temperature $T$. At positive frequencies, we remark that $\mathcal{S}_{11,\text{QPC}}^h(\omega)\approx\mathcal{S}_{22,\text{QPC}}^h(\omega)$ and $ \mathcal{S}_{i\ne j,\text{QPC}}^h(\omega)\approx - \mathcal{S}_{ii,\text{QPC}}^h(\omega)$. A step-like behavior is observed at $\hbar\omega=eV/2$ with the cancellation of both auto-correlators and cross-correlators when $\hbar\omega>eV/2$. This indicates that energy can no longer be transmitted in this case. At negative frequencies, auto-correlators and cross-correlators differ, and all become of negative sign at large negative frequencies (not shown).

\begin{figure}[!h]
	\begin{center}
		\includegraphics[height=3.8cm]{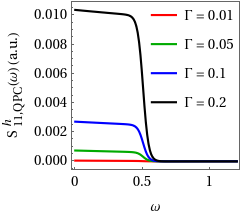}
		\includegraphics[height=3.8cm]{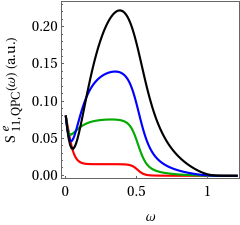}
		\includegraphics[height=3.8cm]{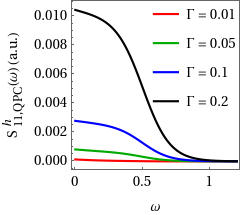}
		\includegraphics[height=3.8cm]{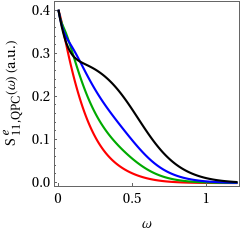}
		\caption{Heat noise $ \mathcal{S}_{11,\text{QPC}}^h(\omega)$ and electrical noise $ \mathcal{S}_{11,\text{QPC}}^e(\omega)$ at positive frequency for various values of $\Gamma$. We set $eV=1$. The temperature is $k_BT=0.02$ for the top plots and $k_BT=0.1$ for the bottom ones.}\label{fig:ShQPCSeQPC}
	\end{center}
\end{figure}
\begin{figure}[!h]
	\begin{center}
		\includegraphics[height=3.7cm]{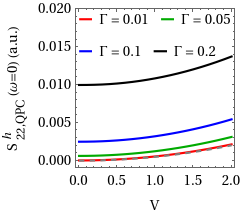}
		\includegraphics[height=3.6cm]{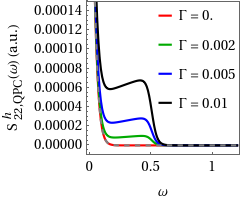}
		\caption{(Left) Zero-frequency heat noise $ \mathcal{S}_{22,\text{QPC}}^h(\omega=0)$ as a function of voltage~$V$, and (Right) Finite-frequency heat noise $ \mathcal{S}_{22,\text{QPC}}^h(\omega)$ as a function of frequency, for various values of $\Gamma$. We set $k_BT=0.02$. The dashed gray lines correspond to the WBS limits of Table \ref{tab:limit_SQOC} at  $\omega=0$ and $T=0$ respectively.} \label{fig:ShQPCSeQPCZF}
	\end{center}
\end{figure}
\begin{figure}[!h]
	\begin{center}
		\includegraphics[height=3.8cm]{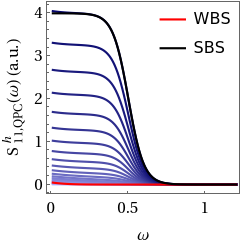}
		\includegraphics[height=3.8cm]{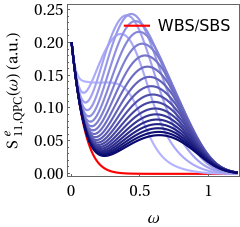}
		\includegraphics[height=3.8cm]{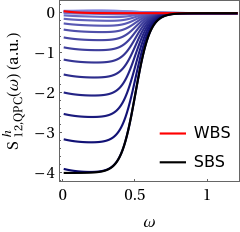}
		\includegraphics[height=3.8cm]{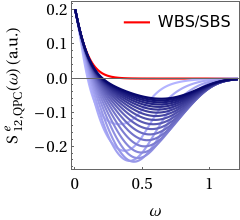}
		\caption{Heat noise and electrical noise at positive frequency from WBS to SBS regimes. We set $eV=1$ and $k_BT=0.05$.}\label{fig:WBStoSBS}
	\end{center}
\end{figure}

In Figs.~\ref{fig:ShQPCSeQPC}, \ref{fig:ShQPCSeQPCZF} and \ref{fig:WBStoSBS}  we focus on the positive frequency region (emission noise). In Fig.~\ref{fig:ShQPCSeQPC}, we plot the heat and electrical noise auto-correlators $ \mathcal{S}_{11,\text{QPC}}^h(\omega)$ and $ \mathcal{S}_{11,\text{QPC}}^e(\omega)$. They both vanish for $\hbar\omega > eV$. This feature reflects the fact that the system cannot emit more energy than it absorbs from its environment. At zero-frequency, we note that the value of the heat noise depends on~$\Gamma$, whereas the electrical noise takes the constant value $4k_BTG_q$, which corresponds to the thermal noise contribution. When the temperature increases, the step-like behavior in the noises at $\hbar\omega=eV/2$ disappears since the thermal noise becomes the dominant contribution in that case. The evolution of the zero frequency heat auto-correlator $\mathcal{S}_{22,\text{QPC}}^h(\omega)$ as a function of voltage and its evolution at finite frequency are displayed in Fig.~\ref{fig:ShQPCSeQPCZF}. At zero frequency, the heat auto-correlator $\mathcal{S}_{22,\text{QPC}}^h(\omega=0)$ follows a power law with voltage $V$ at low $\Gamma$ value (see the red curve in the left plot). Indeed, from Table \ref{tab:limit_SQOC}, we have $ \mathcal{S}_{22,\text{QPC}}^h(\omega=0)= k_BTe^2V^2/(4\pi^2\hbar)$ in the WBS limit. The curves corresponding to the expressions given in Table \ref{tab:limit_SQOC} are plotted with gray dashed lines in  Fig.~\ref{fig:ShQPCSeQPCZF}. Regarding the heat auto-correlator at zero-frequency,  $ \mathcal{S}_{11,\text{QPC}}^h(\omega=0)$, it cancels in the WBS limit due to the $\delta_{2i}\delta_{2j}$ factor (see the red curves in the left plots of Fig.~\ref{fig:ShQPCSeQPC}). The electrical noise at zero frequency does not depend on $\Gamma$, as already underlined, nor on voltage, since we have $ \mathcal{S}_{\text{QPC}}^e(\omega=0)=4k_BTG_q$ (see Table~\ref{tab:limit_SQOC}).

Finally, we discuss the evolution of the NSFFH noise from the WBS regime to the SBS regime. Figure~\ref{fig:WBStoSBS} shows the QPC heat and electrical noises at positive frequency for varying scattering amplitude $\Gamma$. From Table \ref{tab:limit} and Eq.~(\ref{QPCnoise}), we obtain the following results: in the WBS limit, the heat noise is equal to $\mathcal{S}_{ij,\text{QPC}}^h(\omega)=4 \mathcal{\widetilde S}_{ij}^{0}(\omega)$, and in the SBS limit it is equal to $\mathcal{S}_{ij,\text{QPC}}^h(\omega)=(2\delta_{ij}-1)\omega_F\Gamma^2f_-(\omega)/4$ (see also Table~\ref{tab:limit_SQOC}). These two limits are displayed by the black and red curves on the left plots of Fig.~\ref{fig:WBStoSBS}. For what concerns the electric noise, we have $\mathcal{S}_{\text{QPC}}^e(\omega)=4\hbar\omega N(\omega)G_q$ in both the WBS and SBS limits (see the red color curve on the right plots of Fig.~\ref{fig:WBStoSBS}). These results allow us to conclude that the QPC heat noise is a pure partition noise in the SBS limit, whereas the QPC electrical noise is dominated by the Johnson-Nyquist noise contribution.

\section{Conclusion}

In this work, we have investigated the thermal transport properties of a fractional quantum Hall edge state with filling factor $\nu=1/3$ coupled to a $\nu=1$ edge state through a constriction. The system is modeled as an interacting Tomonaga–Luttinger liquid with a localized impurity, characterized by an interaction parameter $\kappa=1/2$, which allows us to employ the refermionization technique and thereby obtain analytical results for the heat current, heat conductance, Lorenz number, and finite-frequency heat noise.

In the large voltage and WBS regime, the Lorenz number approaches $4L_0$, where $L_0=\pi^2 k_B^2/(3e^2)$ is the conventional Lorenz number. This result indicates a rescaling of the Wiedemann–Franz law, which predicts the universal ratio $L_0$ between thermal and electrical conductivities in conventional conductors. The enhancement of the Lorenz number, related to the quasi-particle charge $e^*=e/2$, highlights the nontrivial nature of heat transport in strongly one dimensional correlated systems.

The exact expression for the NSFFH noise, including the heat current auto-correlators and cross-correlators, is given and calculated analytically in various limits, such as zero frequency, zero temperature, WBS or SBS limits. This allows us to identify the regimes in which the dominant contribution comes from thermal noise or from partition noise, and to demonstrate that the thermal transport properties of fractional quantum Hall edge states can depart significantly from conventional metallic behavior. Our combined analysis provides a comprehensive framework for understanding thermal transport and energy fluctuations in interacting quantum conductors. These findings may motivate further theoretical and experimental investigations of NSFFH noise in fractional quantum Hall devices, in particular regarding the interplay between nonlinear transport, impurity scattering, and deviations from conventional transport laws.


\appendix


\section{Heat current calculation}\label{app1}

In this appendix, we present the detailed calculation of the average heat density $\langle\widetilde\rho_{i-}^\text{\,in}(x,\tau) \rangle$ defined in terms of the Hamiltonian $\mathcal{H}_-^\text{in/out}$ given by Eq.~\eqref{hamilton3}, and of the particle number density operator $\mathcal{N}_-^\text{in/out}$ through
\begin{equation}
	\langle\widetilde\rho_{i-}^\text{\,in/out} (x,\tau)\rangle =\langle\mathcal{H}_-^\text{in/out}(x,\tau)\rangle-\mu_i \langle\mathcal{N}_-^\text{in/out}(x,\tau)\rangle.
\end{equation}


First, we begin with the incoming heat density $\langle\rho_{i-}^\text{in}(x,\tau) \rangle$. By substituting the expressions for the Hamiltonian density and the particle number density operator, we obtain
\begin{align}
	\langle\widetilde\rho_{i-}^\text{\,in}(x,\tau) \rangle =&\langle\psi^{\dagger\text{in}}_{-}(x,\tau)\left(-iv_F\frac{\partial}{\partial x}-\omega_0\right)\psi^\text{in}_{-}(x,\tau)\rangle\notag\\
	-&\mu_i \langle\psi^{\dagger\text{in}}_{-}(x,\tau)\psi^\text{in}_{-}(x,\tau)\rangle.
\end{align}
We use the expressions of the field $\psi_-^\text{in}$ given by Eqs.~\eqref{eq:psi1}, and apply the Wick theorem. Thus
\begin{align}
	\langle\widetilde\rho_{i-}^\text{\,in} (x,\tau) \rangle=& \frac{1}{2\pi v_F}\iint_{-\infty}^{+\infty}  d\omega d\omega'\notag\\
	&\times(\omega'-\mu_{i}) e^{i(\omega-\omega')(\tau-\frac{x}{v_F})}\langle a^{\dagger}_{\omega}a_{\omega'}\rangle.
\end{align}
The expectation values are
\begin{align}
\langle a^{\dagger}_{\omega} a_{\omega'} \rangle &= \delta_{\omega,\omega'} \, f_-(\omega),\notag\\
\langle  a_{\omega}a^{\dagger}_{\omega'} \rangle &= \delta_{\omega,\omega'} \big[1- f_-(\omega)\big],
\end{align}
where $f_-(\omega)=(1+\exp((\omega-eV/2)/k_BT))^{-1}$. The final expression for the incoming heat density is
\begin{align}
	\langle\widetilde\rho_{i-}^\text{\,in}  (x,\tau)\rangle =& \frac{1}{2\pi v_F}\int_{-\infty}^{+\infty}  d\omega(\omega-\mu_{i})\big[f_-(\omega)-f_0(\omega)\big].
\end{align}
In the same way, we compute the outgoing heat density
\begin{align}
	&\langle\rho_{i-}^\text{\,out}(x,\tau) \rangle =\bigg\langle\psi^{\dagger \text{out}}_{-}(x,\tau)\left(-iv_F\frac{\partial}{\partial x}-\omega_0\right)\psi^\text{\,out}_{-}(x,\tau)\notag\\
	&+\sqrt{2\pi}\delta(x)\bigg(\Gamma \psi^\text{\,out}_{-}(x,\tau)\mathrm{f}+\Gamma^*\mathrm{f} \psi^{\dagger\text{\,out}}_{-}(x,\tau)\bigg)\bigg\rangle\notag\\
	&-\mu_i \langle\psi^{\dagger \,\text{out}}_{-}(x,\tau)\psi^\text{\,out}_{-}(x,\tau)\rangle.
\end{align}
The second line vanishes because the expectation value of a single operator $\psi^\alpha_{-}$ is zero. It leads to
\begin{align}
	\langle\widetilde\rho_{i-}^\text{\,out} (x,\tau)\rangle &= \frac{1}{2\pi v_F}\iint_{-\infty}^{+\infty}  d\omega d\omega'\notag\\
	&\times(\omega'-\mu_{i}) e^{i(\omega-\omega')(\tau-\frac{x}{v_F})}\langle b^{\dagger}_{\omega}b_{\omega'}\rangle,
\end{align}
where the operator $b_\omega$ is given by $b_\omega=t(\omega)a_\omega+r(\omega)a^\dag_{-\omega}$. Thus
 \begin{align}
	&\langle\widetilde\rho_{i-}^{\,\text{out}}  (x,\tau)\rangle = \frac{1}{2\pi v_F}\iint_{-\infty}^{+\infty}  d\omega \, d\omega' (\omega'-\mu_{i}) \, e^{i(\omega-\omega')(\tau-\frac{x}{v_F})}\notag\\
	& \times\langle  (t^*(\omega) a^\dagger_\omega + r^*(\omega) a_{-\omega}) (t(\omega') a_{\omega'} + r(\omega') a^\dagger_{-\omega'})  \rangle \notag \\
	&= \frac{1}{2\pi v_F}\iint_{-\infty}^{+\infty} d\omega \, d\omega'  (\omega'-\mu_{i}) \, e^{i(\omega-\omega')(\tau-\frac{x}{v_F})}  \notag\\
	&\times \big[ t^*(\omega)t(\omega') \langle  a^\dagger_\omega a_{\omega'}  \rangle + r^*(\omega)r(\omega') \langle  a_{-\omega} a^\dagger_{-\omega'}  \rangle \big].
\end{align}
By applying the Wick theorem to evaluate the expectation values, we obtain
\begin{align}
	\langle\widetilde \rho_{i-}^{\,\text{out}} (x,\tau) \rangle &= \frac{1}{2\pi v_F} \int_{-\infty}^{+\infty} d\omega(\omega - \mu_i ) \notag\\
	&\times \left[ \mathcal{T}(\omega)\left(f_-(\omega)- f_0(\omega)\right) \right.\notag\\
	&+ \left. (1 - \mathcal{T}(\omega))\left(f_+(\omega)- f_0(\omega)\right) \right],
	\label{eq:rho_out_final}
\end{align}
where $\mathcal{T}(\omega)=|t(\omega)|^2 $ is the transmission coefficient obtained from the scattering of the $\psi^\text{out}_-$ mode.


\section{Heat noise calculation}\label{app2}

\subsection{Calculation of $\mathcal{ \widetilde  S}_{i+j+}^{\alpha/\beta}(\omega)$}

The correlator associated to the free field $\psi_+$ is
\begin{eqnarray}
\mathcal{ \widetilde  S}_{i+j+}^{\alpha/\beta}(x,x',\omega)=\frac{v_F^2}{4}\int_{-\infty}^{+\infty} d\tau e^{i\omega \tau} \langle \delta \widetilde\rho_{i+}^{\,\alpha}(x,0) \delta \widetilde\rho_{j+}^{\,\beta}(x',\tau) \rangle.\notag\\
 \end{eqnarray}
Since $\psi_+$ is a free field, the result will not depend on $\alpha$, nor on $\beta$. The calculation yields
\begin{eqnarray}
 &&\langle \widetilde\rho_{i+}^{\,\alpha}(x,0)  \widetilde\rho_{j+}^{\,\beta}(x',\tau) \rangle
 =\frac{1}{4\pi^2v_F^2}\int_{-\infty}^{+\infty} d\omega_1 d\omega_2 d\omega_3 d\omega_4\notag\\
 &&\times(\omega_2-\mu_i)(\omega_4-\mu_j)e^{i(\omega_3-\omega_4)\tau}
 \langle a^\dagger_{\omega_1}a_{\omega_2}a^\dagger_{\omega_3}a_{\omega_4}\rangle.
\end{eqnarray}
By applying the Wick theorem, and by subtracting the $\langle \widetilde\rho_{i+}^{\,\alpha} \rangle \langle\widetilde\rho_{j+}^{\,\beta}\rangle$ term, we obtain
\begin{eqnarray}
\mathcal{ \widetilde  S}_{i+j+}^{\alpha/\beta}(\omega)&=&\frac{1}{16\pi^2}\int_{-\infty}^{+\infty} d\omega_1 d\omega_2 d\omega_3 d\omega_4{ d\tau  e^{i(\omega+\omega_3-\omega_4)\tau}} \notag\\
 &&\times(\omega_2-\mu_i)(\omega_4-\mu_j)
 {\langle a^\dagger_{\omega_1}a_{\omega_4}\rangle}
 {\langle  a_{\omega_2}a^\dagger_{\omega_3}\rangle},
 \end{eqnarray}
which leads to
 \begin{eqnarray}\label{Cpp}
\mathcal{ \widetilde  S}_{i+j+}^{\alpha/\beta}(\omega)= \mathcal{ \widetilde  S}_{ij}^{0}(\omega),
 \end{eqnarray}
 where we have defined
  \begin{eqnarray}\label{defS0}
 \mathcal{ \widetilde  S}_{ij}^{0}(\omega)&=&\frac{1}{16\pi^2}\int_{-\infty}^{+\infty} d\omega'(\omega'-\omega-\mu_i)(\omega'-\mu_j)\notag\\
&&\times f_-(\omega')(1-f_-(\omega'-\omega)).
 \end{eqnarray}

\subsection{Calculation of $\mathcal{ \widetilde  S}_{i-j-}^{\alpha/\beta}(\omega)$}

The correlator associated to the field $\psi_-$ is
\begin{eqnarray}
\mathcal{ \widetilde  S}_{i-j-}^{\alpha/\beta}(\omega)=\frac{v_F^2}{4}\int_{-\infty}^{+\infty} d\tau e^{i\omega \tau} \langle \delta \widetilde\rho_{i-}^{\,\alpha}(x,0) \delta \widetilde\rho_{j-}^{\,\beta}(x',\tau) \rangle . \notag\\
 \end{eqnarray}
It depends on $\alpha$ and $\beta$. In the following, we calculate each term separately.

\subsubsection{Calculation of $\mathcal{ \widetilde  S}_{i-j-}^{\text{in}/\text{in}}(\omega)$}

The field $\psi^\text{in}_{-}$ does not feel the scattering, so it can be considered as a free field, as $\psi^\alpha_{+}$. In consequence, we have
 \begin{eqnarray}
\mathcal{ \widetilde  S}_{i-j-}^{\,\text{in}/\text{in}}(\omega)=\mathcal{ \widetilde  S}_{ij}^{0}(\omega),
 \end{eqnarray}
where $\mathcal{ \widetilde  S}_{ij}^{0}(\omega)$ is defined in Eq.~(\ref{defS0}).

\subsubsection{Calculation of $\mathcal{ \widetilde  S}_{i-j-}^{\text{out}/\text{in}}(\omega)$}

We have to calculate
 \begin{eqnarray}
 &&\langle \widetilde\rho_{i-}^{\,\text{out}}(x,0)  \widetilde\rho_{j-}^{\,\text{in}}(x',\tau) \rangle
 =\frac{1}{4\pi^2v_F^2}\int_{-\infty}^{+\infty} d\omega_1 d\omega_2 d\omega_3 d\omega_4\notag\\
 &&\times(\omega_2-\mu_i)(\omega_4-\mu_j)e^{i(\omega_3-\omega_4)\tau}
 \langle b^\dagger_{\omega_1}b_{\omega_2}a^\dagger_{\omega_3}a_{\omega_4}\rangle.
\end{eqnarray}
It leads to
 \begin{eqnarray}
 &&\mathcal{ \widetilde  S}_{i-j-}^{\,\text{out}/\text{in}}(\omega)
 =\frac{1}{16\pi^2}\int_{-\infty}^{+\infty} d\omega_1 d\omega_2 d\omega_3 d\omega_4d\tau \notag\\
 &&\times e^{i\omega \tau} e^{i(\omega_3-\omega_4)\tau}(\omega_2-\mu_i)(\omega_4-\mu_j)\notag\\
 &&\times\left(\langle b^\dagger_{\omega_1}a_{\omega_4}\rangle\langle b_{\omega_2}a^\dagger_{\omega_3}\rangle
 -\langle b^\dagger_{\omega_1}a^\dagger_{\omega_3}\rangle\langle b_{\omega_2}a_{\omega_4}\rangle  \right),
\end{eqnarray}
that is
 \begin{eqnarray}
 &&\mathcal{ \widetilde  S}_{i-j-}^{\,\text{out}/\text{in}}(\omega)
=\frac{1}{16\pi^2}\int_{-\infty}^{+\infty}  d\omega_1 d\omega_2 d\omega_3 d\omega_4 d\tau  e^{i(\omega+\omega_3-\omega_4)\tau}\notag\\
 && \times   (\omega_2-\mu_i)(\omega_4-\mu_j)
 \big(t^*(\omega_1) t(\omega_2)\langle a^\dagger_{\omega_1}a_{\omega_4}\rangle
 \langle a_{\omega_2}a^\dagger_{\omega_3}\rangle\notag\\
 &&-r(\omega_2)r^*(\omega_1)\langle a_{-\omega_1}a^\dagger_{\omega_3}\rangle\langle a^\dagger_{-\omega_2}a_{\omega_4}\rangle \big).
\end{eqnarray}
By setting $\omega_4=\omega'$, and by using the relations $t(-\omega)=t^*(\omega)$ and $r(-\omega)=r^*(\omega)$, we finally obtain
 \begin{eqnarray}\label{Soutin}
 &&\mathcal{ \widetilde  S}_{i-j-}^{\,\text{out}/\text{in}}(\omega)
 =\frac{1}{16\pi^2}\int_{-\infty}^{+\infty}  d\omega' (\omega'-\mu_j) f_-(\omega')[1-f_-(\omega'-\omega)]\notag\\
 &&\times[(\omega'-\omega-\mu_i)t^*(\omega') t(\omega'-\omega)+(\omega'+\mu_i)r^*(\omega')r(\omega'-\omega)].\notag\\
\end{eqnarray}

\subsubsection{Calculation of $\mathcal{ \widetilde  S}_{i-j-}^{\text{in}/\text{out}}(\omega)$}

The calculation is similar to one of the previous subsection. We obtain $\mathcal{ \widetilde  S}_{i-j-}^{\,\text{in}/\text{out}}(\omega)= (\mathcal{ \widetilde  S}_{i-j-}^{\,\text{out}/\text{in}}(\omega))^*$, that is
 \begin{eqnarray}\label{Sinout}
 &&\mathcal{ \widetilde  S}_{i-j-}^{\,\text{in}/\text{out}}(\omega)
 =\frac{1}{16\pi^2}\int_{-\infty}^{+\infty} d\omega' (\omega'-\mu_j) f_-(\omega')[1-f_-(\omega'-\omega)]\notag\\
 &&\times[(\omega'-\omega-\mu_i)t(\omega') t^*(\omega'-\omega)+(\omega'+\mu_i)r(\omega')r^*(\omega'-\omega)].
\notag\\
\end{eqnarray}

\subsubsection{Calculation of $\mathcal{ \widetilde  S}_{i-j-}^{\text{out}/\text{out}}(\omega)$}

We have to calculate
\begin{eqnarray}
 &&\langle \widetilde\rho_{i-}^{\,\text{out}}(x,0)  \widetilde\rho_{j-}^{\,\text{out}}(x',\tau) \rangle
 =\frac{1}{4\pi^2v_F^2}\int_{-\infty}^{+\infty} d\omega_1 d\omega_2 d\omega_3 d\omega_4 \notag\\
 &&\times(\omega_2-\mu_i)(\omega_4-\mu_j)e^{i(\omega_3-\omega_4)\tau}
 \langle b^\dagger_{\omega_1}b_{\omega_2}b^\dagger_{\omega_3}b_{\omega_4}\rangle\notag\\
 &&+\frac{\omega_F}{v_F^2}\int_{-\infty}^{+\infty} d\omega_1  d\omega_3
 (|\Gamma|^2e^{-i\omega_3\tau}\langle b^\dagger_{\omega_1}b_{\omega_3}\rangle
 +|\Gamma|^2e^{i\omega_3\tau}\langle b_{\omega_1}b^\dagger_{\omega_3}\rangle
\notag\\
 && +\Gamma^2e^{-i\omega_3\tau}\langle b_{\omega_1}b_{\omega_3}\rangle
 +(\Gamma^*)^2e^{i\omega_3\tau}\langle b^\dagger_{\omega_1}b^\dagger_{\omega_3}\rangle).
\end{eqnarray}
It leads to
 \begin{eqnarray}
 \mathcal{ \widetilde  S}_{i-j-}^{\,\text{out}/\text{out}}(\omega)
 &=&\frac{1}{16\pi^2}\int_{-\infty}^{+\infty} d\omega_1 d\omega_2 d\omega_3 d\omega_4 d\tau  e^{i(\omega+\omega_3-\omega_4)\tau} \notag\\
 &&\times (\omega_2-\mu_i)(\omega_4-\mu_j)
 \notag\\
 &&\times \bigg(\langle b^\dagger_{\omega_1}b_{\omega_4}\rangle\langle b_{\omega_2}b^\dagger_{\omega_3}\rangle
 -\langle b^\dagger_{\omega_1}b^\dagger_{\omega_3}\rangle\langle b_{\omega_2}b_{\omega_4}\rangle  \bigg)\notag\\
 &&+\frac{\omega_F}{4}\int_{-\infty}^{+\infty} d\omega_1  d\omega_3 d\tau e^{i(\omega-\omega_3) \tau}
 \notag\\
 &&\times \bigg(|\Gamma|^2\langle b^\dagger_{\omega_1}b_{\omega_3}\rangle
 + \Gamma^2 \langle b_{\omega_1}b_{\omega_3}\rangle\bigg)\notag\\
 &&+\frac{\omega_F}{4}\int_{-\infty}^{+\infty} d\omega_1  d\omega_3 d\tau e^{i(\omega+\omega_3)\tau}
 \notag\\
 &&\times \bigg(|\Gamma|^2 \langle b_{\omega_1}b^\dagger_{\omega_3}\rangle
+(\Gamma^*)^2 \langle b^\dagger_{\omega_1}b^\dagger_{\omega_3}\rangle\bigg).
\end{eqnarray}
By using the relations
\begin{eqnarray}
\langle b^{\dagger}_{\omega} b_{\omega'} \rangle &=& \delta_{\omega,\omega'}[\mathcal{T}(\omega) f_-(\omega) + \mathcal{R}(\omega) f_+(\omega) ],\notag  \\
\langle  b_{\omega}b^{\dagger}_{\omega'} \rangle &=&\delta_{\omega,\omega'} [\mathcal{T}(\omega)(1- f_-(\omega)) + \mathcal{R}(\omega)(1- f_+(\omega))],\notag \\
\langle b^{\dagger}_{\omega} b^{\dagger}_{\omega'} \rangle &=& \delta_{\omega,-\omega'} [t^*(\omega)r(\omega) f_-(\omega) + t(\omega)r^*(\omega) f_+(\omega) ], \notag\\
\langle  b_{\omega}b_{\omega'} \rangle& =&\delta_{\omega,-\omega'} [ t(\omega)r^*(\omega)(1- f_-(\omega))\notag\\
&&+t^*(\omega)r(\omega)(1- f_+(\omega))],
\end{eqnarray}
we obtain
 \begin{eqnarray}
 \mathcal{ \widetilde  S}_{i-j-}^{\,\text{out}/\text{out}}(\omega)
 &=&\frac{1}{16\pi^2}\int_{-\infty}^{+\infty} d\omega'(\omega'-\omega-\mu_i)(\omega'-\mu_j)
\notag\\
 &&\times[ \mathcal{T}(\omega') f_-(\omega') + \mathcal{R}(\omega') f_+(\omega') ]\notag\\
 &&\times  [ \mathcal{T}(\omega'-\omega)(1- f_-(\omega'-\omega)) \notag\\
 &&+ \mathcal{R}(\omega'-\omega)(1- f_+(\omega'-\omega))]\notag\\
 &&+\frac{1}{16\pi^2}\int_{-\infty}^{+\infty} d\omega'(\omega'+\mu_i)(\omega'-\mu_j) \notag\\
 &&\times [t^*(\omega-\omega')r(\omega-\omega') f_-(\omega-\omega')\notag\\
 && + t(\omega-\omega')r^*(\omega-\omega') f_+(\omega-\omega') ]\notag\\
 &&\times
 [ t(-\omega')r^*(-\omega')(1- f_-(-\omega'))\notag\\
 && +t^*(-\omega')r(-\omega')(1- f_+(-\omega'))] \notag\\
 &&+\frac{\omega_F}{4}
 \Big(|\Gamma|^2 [ \mathcal{T}(\omega) f_-(\omega) + \mathcal{R}(\omega) f_+(\omega) ]
 \notag\\
 &&+|\Gamma|^2[ \mathcal{T}(\omega)f_+(\omega) + \mathcal{R}(\omega)f_-(\omega)]\Big)\notag\\
 &&+\frac{\omega_F}{4}
 \Big(\Gamma^2 [ t^*(\omega)r(\omega) f_+(\omega) +t(\omega)r^*(\omega)f_-(\omega)]
 \notag\\
 &&+(\Gamma^*)^2[t^*(\omega)r(\omega) f_-(\omega) + t(\omega)r^*(\omega) f_+(\omega) ]\Big).\notag\\
\end{eqnarray}
By using the relations $1-f_\mp(-\omega)=f_\pm(\omega)$, $r(\omega)=1-t(\omega)$, $\mathcal{T}(\omega)=|t(\omega)|^2$, $t(\omega)-t^*(\omega)=2t(\omega)r^*(\omega)$ and $t(\omega)+t^*(\omega)=2\mathcal{T}(\omega)$, the result simplifies to
 \begin{eqnarray}\label{Soutout}
 \mathcal{ \widetilde  S}_{i-j-}^{\,\text{out}/\text{out}}(\omega)
&=&\frac{1}{16\pi^2}\int_{-\infty}^{+\infty} d\omega'(\omega'-\omega-\mu_i)(\omega'-\mu_j)\notag\\
&&\times[ \mathcal{T}(\omega') f_-(\omega') +(1- \mathcal{T}(\omega')) f_+(\omega') ]\notag\\
 &&\times  [ \mathcal{T}(\omega'-\omega)(1- f_-(\omega'-\omega)) \notag\\
&&+(1- \mathcal{T}(\omega'-\omega))(1- f_+(\omega'-\omega))]\notag\\
 &&+\frac{1}{16\pi^2}\int_{-\infty}^{+\infty} d\omega'(\omega'+\mu_i)(\omega'-\mu_j)\notag\\
&&\times t(\omega')r^*(\omega')t(\omega'-\omega)r^*(\omega'-\omega) \notag\\
 &&\times [f_+(\omega')-f_-(\omega')][f_+(\omega'-\omega)- f_-(\omega'-\omega)]
 \notag\\
&&+\frac{\omega_F\Gamma^2}{4} ( f_-(\omega) +f_+(\omega)).
\end{eqnarray}


\section{Calculation of some limit cases}\label{app3}

The expressions for the heat noise in the limit cases are calculated starting from Eqs.~(\ref{expSinin}) to (\ref{expSoutout}). We have $\mathcal{ \widetilde  S}_{i-j-}^{\,\text{in}/\text{in}}(\omega)=\mathcal{ \widetilde  S}_{ij}^{0}(\omega)$, and we calculate $\mathcal{ \widetilde  S}_{i-j-}^{\,\text{out}/\text{in}}(\omega)$ and $\mathcal{ \widetilde  S}_{i-j-}^{\,\text{out}/\text{out}}(\omega)$ in the various limits below.

{\it Zero-frequency limit --} By setting $\omega=0$ in Eqs.~(\ref{expSinin}) to~(\ref{expSoutout}), we obtain

 \begin{eqnarray}
  \mathcal{ \widetilde  S}_{i-j-}^{\,\text{in}/\text{in}}(0)&=&\mathcal{ \widetilde  S}_{ij}^{0}(0)\notag\\
&=&\frac{1}{16\pi^2}\int_{-\infty}^{+\infty} d\omega'(\omega'-\mu_i)(\omega'-\mu_j) \notag\\
 &&\times f_-(\omega')[1-f_-(\omega')],
 \end{eqnarray}

 \begin{eqnarray}
 \mathcal{ \widetilde  S}_{i-j-}^{\,\text{out}/\text{in}}(0)
 &=&\frac{1}{16\pi^2}\int_{-\infty}^{+\infty}  d\omega' (\omega'-\mu_j)\notag\\
 &&\times[(\omega'-\mu_i)\mathcal{T}(\omega')+(\omega'+\mu_i)\mathcal{R}(\omega')] \notag\\
 &&\times f_-(\omega')[1-f_-(\omega')],
\end{eqnarray}
where $\mathcal{R}(\omega')=1-\mathcal{T}(\omega')$, and
 \begin{eqnarray}
 \mathcal{ \widetilde  S}_{i-j-}^{\,\text{out}/\text{out}}(0)
&=&\frac{1}{16\pi^2}\int_{-\infty}^{+\infty} d\omega'(\omega'-\mu_i)(\omega'-\mu_j)\notag\\
&&\times[ \mathcal{T}(\omega') f_-(\omega') +(1- \mathcal{T}(\omega')) f_+(\omega') ]\notag\\
 &&\times  [ \mathcal{T}(\omega')(1- f_-(\omega')) \notag\\
&&+(1- \mathcal{T}(\omega'))(1- f_+(\omega'))]\notag\\
 &&+\frac{1}{16\pi^2}\int_{-\infty}^{+\infty} d\omega'(\omega'+\mu_i)(\omega'-\mu_j)\notag\\
&&\times [t(\omega')r^*(\omega')]^2 [f_+(\omega')-f_-(\omega')]^2
 \notag\\
&&+\frac{\omega_F\Gamma^2}{4}(f_-(0)+f_+(0)).
\end{eqnarray}

{\it At equilibrium and zero-frequency --} For $V=0$ and $\omega=0$,  Eqs.~(\ref{expSinin}) to~(\ref{expSoutout}) lead to
\begin{eqnarray}
\mathcal{ \widetilde  S}_{i-j-}^{\,\text{in}/\text{in}}(0)= \mathcal{ \widetilde  S}_{i-j-}^{\,\text{out}/\text{in}}(0)=\mathcal{ \widetilde  S}_{ij}^{0}(0)\Big|_{V=0}
\end{eqnarray}
where
 \begin{eqnarray}
\mathcal{ \widetilde  S}_{ij}^{0}(0)\Big|_{V=0}
=\frac{1}{16\pi^2}\int_{-\infty}^{+\infty} d\omega'\omega'^2  f_0(\omega')[1-f_0(\omega')],\notag\\
 \end{eqnarray}
and
 \begin{eqnarray}
 \mathcal{ \widetilde  S}_{i-j-}^{\,\text{out}/\text{out}}(0)
&=&\mathcal{ \widetilde  S}_{ij}^{0}(0)\Big|_{V=0}+\frac{\omega_F\Gamma^2}{2}f_0(0).
\end{eqnarray}

{\it Zero-temperature limit --}  For $T=0$ and $\omega\ge 0$, Eqs.~(\ref{expSinin}) to~(\ref{expSoutout}) give $\mathcal{ \widetilde  S}_{i-j-}^{\,\text{in}/\text{in}}(\omega)=\mathcal{ \widetilde  S}_{i-j-}^{\,\text{out}/\text{in}}(\omega)=0$ since $f_-(\omega')[1-f_-(\omega'-\omega)]=f_+(\omega')[1-f_\pm(\omega'-\omega)]=0$, and
 \begin{eqnarray}
 \mathcal{ \widetilde  S}_{i-j-}^{\,\text{out}/\text{out}}(\omega)
&=&\frac{1}{16\pi^2}\int_{\omega-eV/2}^{+eV/2}  d\omega'(\omega'-\mu_j)\notag\\
&&\times \Big[(\omega'-\omega-\mu_i)\mathcal{T}(\omega')     (1- \mathcal{T}(\omega'-\omega))\notag\\
 &&+(\omega'+\mu_i) t(\omega')r^*(\omega')t(\omega'-\omega)r^*(\omega'-\omega)\Big]
 \notag\\
&&+\frac{\omega_F\Gamma^2}{4}f_-(\omega),
\end{eqnarray}
since $f_-(\omega')[1-f_+(\omega'-\omega)]=1$ if $\omega'\in[\omega-eV/2,eV/2]$, $f_-(\omega')-f_+(\omega')=1$ if $\omega'\in[-eV/2,eV/2]$ and both are equal to zero otherwise. In conclusion, at $T=0$, there is no thermal noise, and no over-bias noise.

{\it WBS limit --} When $\Gamma\rightarrow 0$, we have $t(\omega')\approx 1$ and $r(\omega')\approx 0$, thus Eqs.~(\ref{expSinin}) to~(\ref{expSoutout}) give $\mathcal{ \widetilde  S}_{i-j-}^{\,\alpha/\beta}(\omega)=\mathcal{ \widetilde  S}_{ij}^{0}(\omega)$, $\forall \alpha$ and $\beta$, meaning that only thermal noise survives.

{\it SBS limit --}  When $\Gamma$ is large, we have $t(\omega')\approx 0$ and $r(\omega')\approx 1$, thus Eqs.~(\ref{expSinin}) to~(\ref{expSoutout}) give $\mathcal{ \widetilde  S}_{i-j-}^{\,\text{in}/\text{in}}(\omega)=\mathcal{ \widetilde  S}_{ij}^{0}(\omega)$, and
 \begin{eqnarray}
 \mathcal{ \widetilde  S}_{i-j-}^{\,\text{out}/\text{in}}(\omega)
 &=&\frac{1}{16\pi^2}\int_{-\infty}^{+\infty}  d\omega' (\omega'-\mu_j)(\omega'+\mu_i) \notag\\
 &&\times f_-(\omega')[1-f_-(\omega'-\omega)],
\end{eqnarray}
 and
 \begin{eqnarray}
 \mathcal{ \widetilde  S}_{i-j-}^{\,\text{out}/\text{out}}(\omega)
&=&\frac{1}{16\pi^2}\int_{-\infty}^{+\infty} d\omega'(\omega'-\omega-\mu_i)(\omega'-\mu_j)\notag\\
&&\times f_+(\omega') (1- f_+(\omega'-\omega))\notag\\
&&+\frac{\omega_F\Gamma^2}{4} ( f_-(\omega) +f_+(\omega)).
\end{eqnarray}
At zero-temperature $T=0$ and positive frequency $\omega\ge 0$, we have $f_-(\omega')[1-f_-(\omega'-\omega)]=f_+(\omega')[1-f_+(\omega'-\omega)]=f_+(\omega)=0$, thus $\mathcal{ \widetilde  S}_{i-j-}^{\,\text{in}/\text{in}}(\omega)=\mathcal{ \widetilde  S}_{i-j-}^{\,\text{out}/\text{in}}(\omega)=0$ and
 \begin{eqnarray}
 \mathcal{ \widetilde  S}_{i-j-}^{\,\text{out}/\text{out}}(\omega)
=\frac{\omega_F\Gamma^2}{4} f_-(\omega) .
\end{eqnarray}

\bibliography{ref}

\end{document}